\documentclass[aps,prb,showpacs,reprint,superscriptaddress]{revtex4-1}

\usepackage{graphicx}
\usepackage{color}
\usepackage{amsmath}
\usepackage{bbm}
\usepackage{amssymb}

\usepackage{dsfont}

\usepackage[utf8]{inputenc}	% kodowanie UTF-8
\usepackage[T1]{fontenc}

\input epsf

\begin{document}

\title{Superconductivity and intra-unit-cell electronic nematic phase in the three-band model of cuprates}
\author{M. Zegrodnik}
\email{michal.zegrodnik@agh.edu.pl}
\affiliation{Academic Centre for Materials and Nanotechnology, AGH University of Science and Technology, Al. Mickiewicza 30, 30-059 Krakow,
Poland}
\author{A. Biborski}
\email{andrzej.biborski@agh.edu.pl}
\affiliation{Academic Centre for Materials and Nanotechnology, AGH University of Science and Technology, Al. Mickiewicza 30, 30-059 Krakow,
Poland}
%\author{M. Fidrysiak}
%\email{maciej.fidrysiak@uj.edu.pl}
%\affiliation{Marian Smoluchowski Institute of Physics, 
%Jagiellonian University, ul. \L ojasiewicza 11,
%30-348 Krakow, Poland}
\author{J. Spa\l ek}
\email{jozef.spalek@uj.edu.pl}
\affiliation{Marian Smoluchowski Institute of Physics, 
Jagiellonian University, ul. \L ojasiewicza 11,
30-348 Krakow, Poland}

\date{02.07.2019}

\begin{abstract}
The intra-unit-cell nematic phase is studied within the three-band Emery model of the cuprates with the use of the approach based on the \textit{diagrammatic expansion of the Gutzwiller wave function} (DE-GWF). According to our analysis the spontaneous $C_4$ symmetry breaking of the electronic wave function, leading to the nematic behavior, can appear due to electron correlations induced mainly by the onsite Coulomb repulsion, even in the absence of the corresponding intersite oxygen-oxygen repulsion term. The latter has been considered as the triggering factor of the nematic state formation in a number of previous studies. Also, we show that, at the transition to the nematic phase electron concentration transfer from $d$- to $p$- orbitals takes place, apart from the usually discussed $p_x/p_y$ polarization. The determined stability regime of the nematic phase appears in the doping range similar to that of the paired phase, showing that both phases have a common origin, even though they compete. Also, we show that in a significant doping range a coexistence region of superconductivity and nematicity appears. The results are discussed in the view of the experimental findings considering the relation between nematicity and pseudogap behavior.

\end{abstract}

\maketitle
%%%%%%%%%%%%%%%%%%%%%%%%%%%%%%%%%%%%%%%%%%%%%%%%%%%%%%%%%%%%%%%%%%%%%%%%
\section{Introduction}

A number of broken symmetry states appear in the cuprate high temperature superconductors. One of the key issues is to identify the mechanism of their creation and how they are interrelated. The four-fold ($C_4$) rotational symmetry breaking with the preservation of the translational symmetry, which is observed in the cuprates\cite{Lawler2010} and titanium-oxypnictides\cite{Frandsen2014} corresponds to the creation of the so-called nematic phase. Due to the structural LTT phase transition in La- based cuprates or the orthorombic distortion in YBCO, it is difficult to validate the intrinsic nematic behavior of the electronic wave function, as the $C_4$ symmetry is already broken by the crystal lattice distortion. Nevertheless, it has been argued that a significant contribution to the nematicity is distinct from the effects related to the lattice\cite{Achkar2016,Sato2017}. The STM measurements on Bi-2212 and NCCOC seem to show a more direct evidence of electronic nematicity, which is not related to the structure\cite{Lawler2010,Mesaros2011}. This suggests that one of the generic features of the copper-based compounds may be the intrinsic susceptibility towards the $C_4$ symmetry breaking of the electronic wave function in the CuO$_2$ planes.

It has been established that the nematic ordering in the cuprates arises from the differences in electron concentrations at the two oxygen sites within each unit cell of the copper-oxygen plane \cite{Lawler2010}. An analogus situation takes place in the titanium-based materials\cite{Frandsen2014}. Such a charge shift between the $p_x$/$p_y$ orbitals is also reported in the charge-density wave (CDW) phase of the cuprates\cite{Comin2015}.
 Therefore, the connection between the two phases has been discussed\cite{Cyr2015,Achkar2016,Pelc2016}. In particular, it has been suggested, that the nematicity may be understood as a precursor state before the formation of charge ordering, in which both the $C_4$ and translational symmetries are broken\cite{Pelc2016}. Also, in some analysis the nematic phase has been related to the appearance of the so-called pseudogap phase\cite{Daou2010,Lawler2010,Sato2017}. In fact, a strong thermodynamic evidence for the nematic character of the pseudogap state has been reported recently\cite{Sato2017}. However, the question if the $C_4$ symmetry breaking is the primary cause or a secondary effect of the pseudogap behavior still remains open. Nevertheless, since the pseudogap phenomenon is reported down to $T\approx0$ K and is connected with the $C_4$ symmetry breaking\cite{Lawler2010,Hashimoto2014}, than both superconductivity and nematicity should appear simultaneously in a significant doping range. Again, it is not clear if the pairing appears inside the nematic domains leading to a coexistent superconducting-nematic phase or a phase-separation scenario is realized. 
 
The nematicity has been studied theoretically in the single band models, used for the effective description of the Cu-O planes of the cuprates\cite{Yamase2000, Okamoto2010, Yamase2007, Kaczmarczyk2016, Kitatani2017, Zegrodnik2018, Slizovskiy2018}. Due to the intra-unit-cell character of the nematic phase formation, a more realistic description should include explicitly the oxygen degrees of freedom. Therefore, the three-band Emery model has also been applied with respect to the considered symmetry breaking within the mean field approach\cite{Kivelson2004,Fischer2011} or more sophisticated methods \cite{Bulut2013,Tsuchiizu2018}. In these considerations the Coulomb repulsion between the oxygen orbitals played an important role leading to the nematic instability\cite{Kivelson2004,Fischer2011,Bulut2013} or the spin-fluctuation-driven mechanism has been proposed in the strong coupling limit\cite{Tsuchiizu2018}. 

Here, we analyze the $C_4$ symmetry breaking resulting from the $p$-orbital polarization, $n_{p_x}\neq n_{p_y}$, in the three band Emery model, with the values of the microscopic parameters appropriate for the cuprates. To focus purely on the susceptibility towards the nematic instability of the electronic wave function we consider the ideal square lattice situation. Within our approach the nematicity appears as a result of strong inter-electronic correlations, which are taken into account by the higher order terms of the diagrammatic expansion of the Gutzwiller wave function (DE-GWF method). The method has been recently applied to both the single- and three-band descriptions of the paired phase of the cuprates and leads to good agreement with the principal experimental observations\cite{Zegrodnik_1band_1,Zegrodnik_1band_2,Zegrodnik_3band}. In contrast to the previous results obtained for the Emery model\cite{Kivelson2004,Fischer2011,Bulut2013,Tsuchiizu2018}, we show that the nematic behavior of the electronic wave function can be induced by strong inter-electronic correlations, with the dominant role of the onsite Coulomb repulsion at the copper $d$-orbitals, even without the corresponding intersite oxygen-oxygen term. Such a result is also supported by previous analysis carried out for the single-band Hubbard model\cite{Kitatani2017,Kaczmarczyk2016,Zegrodnik2018}. We also study the interplay between the $d$-$wave$ pairing and nematic phase. In particular, according to our interpretation the $C_4$ symmetry breaking and the paired state seem to have the same origin and in a significant doping range superconductivity and nematic phase coexist (SC+N), in spite of the circumstance that the two compete. This last result is discussed in view of the experimental findings considering the relation between the nematicity and the pseudogap behavior\cite{Sato2017,Hashimoto2014}.

The paper is organized as follows. In the subsequent Section we present the theoretical model and provide some details of the DE-GWF calculation scheme. Next, the results corresponding to the pure nematic phase are discussed, before turning to the analysis of the nematicity-superconductivity interplay. The conclusions are deferred to the last Section.

\section{Theory}
\section{Model and methods}
\begin{figure}
 \centering
 \includegraphics[width=0.30\textwidth]{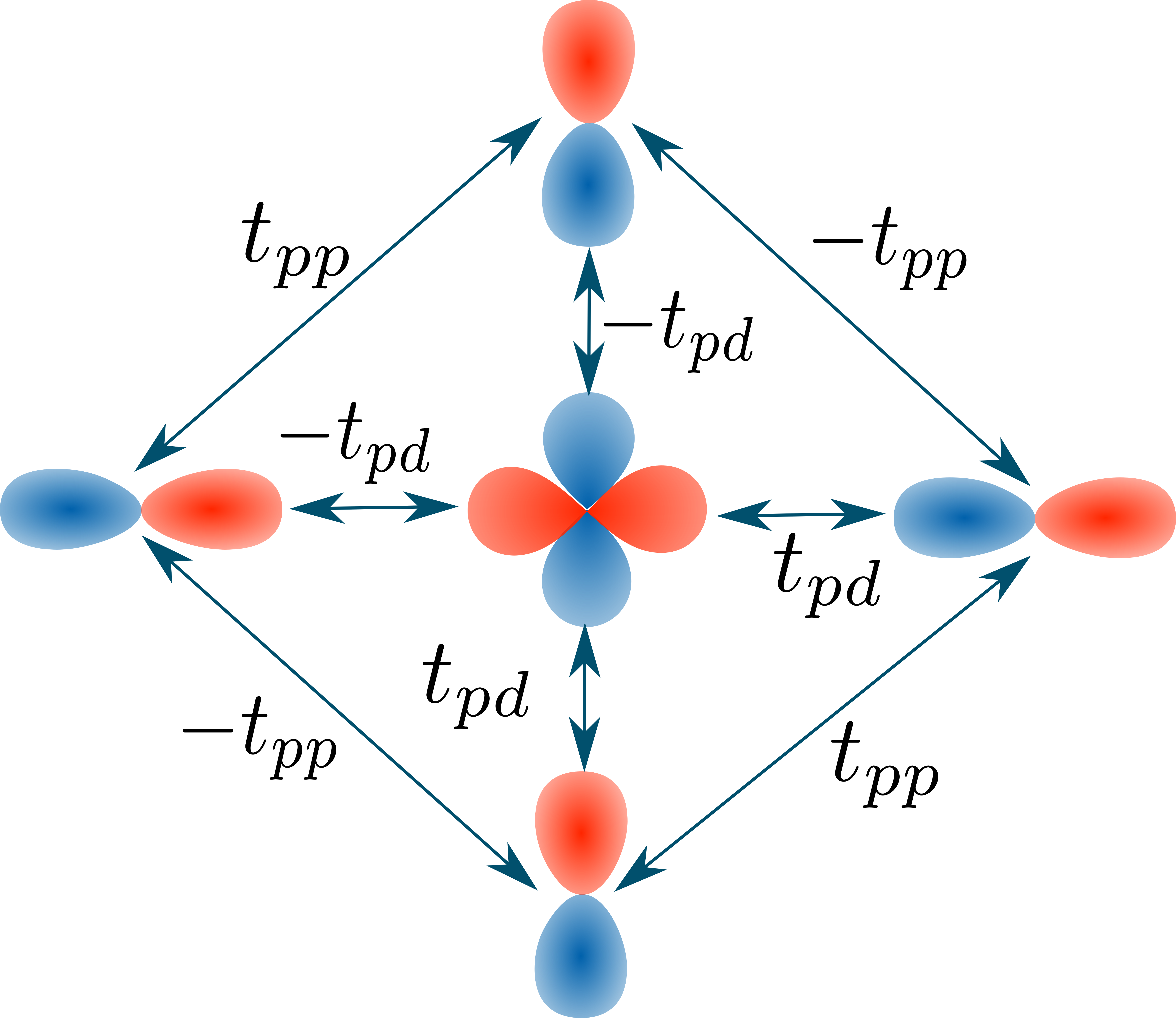}
 \caption{The hopping parameters between the three types of orbitals in the model and the corresponding sign convention for the antibonding orbital structure. The $d_{x^2-y^2}$  orbital is centered at the copper site and the $p_x/p_y$ orbitals are centered at the oxygen sites.}
 \label{fig:Cu_O_hoppings}
\end{figure}

We start from the three-band Emery model in the electron representation of the form
\begin{equation}
\begin{split}
 \hat{H}&=\sum_{\langle il,jl'\rangle}t^{ll'}_{il}\hat{c}^{\dagger}_{il\sigma}\hat{c}_{jl'\sigma}+\sum_{il}(\epsilon_{l}-\mu)\hat{n}_{il}+\sum_{il}U_{l}\hat{n}_{il\uparrow}\hat{n}_{il\downarrow},\\ 
 \end{split}
 \label{eq:Hamiltonian_start}
\end{equation}
where $\hat{c}^{\dagger}_{il\sigma}$ ($\hat{c}_{il\sigma}$) are the creation (anihilation) operators of electrons with spin $\sigma$ at the $i$-th atomic site and orbital denoted by $l\in\{{d,p_x,p_y}\}$. The notation $\langle il,jl'\rangle$ means that the summation is carried out only for the interorbital nearest neighbor hoppings (cf. Fig. \ref{fig:Cu_O_hoppings}). The notation of the corresponding hopping energies is shown in Fig \ref{fig:Cu_O_hoppings}. The $p$ orbitals are located at the oxygen atomic sites which reside in between every two nearest neighbor copper sites (containing the $d$ orbital states) located at the nodes of the square lattice. In such a structure a single unit cell contains one copper and two oxygen atomic sites. The second term of the Hamiltonian defines to the $d$ and $p_x/p_y$ atomic levels ($\epsilon_{p_x}=\epsilon_{p_y}\equiv\epsilon_{p}$, $\epsilon_d-\epsilon_p\equiv\epsilon_{dp}$), together with the chemical potential contribution. The interaction parameters $U_d$ and $U_{p_x}=U_{p_y}\equiv U_p$ represent the intrasite Coulomb repulsions between two electrons with opposite spins located on the $d$ and $p_x/p_y$ orbitals, respectively. 

The model represents an effective description of a single copper-oxygen plane of the cuprates. Here, we take the typical values of the hopping energies $t_{dp}=1.13\;$eV, $t_{pp}=0.49\;$eV, and the charge-transfer energy $\epsilon_{dp}=3.2\;$eV. The commonly used values of the interaction parameter $U_d$ ($U_p$) range between $8-10.5\;$eV ($4-6\;$eV), depending on the particular approach\cite{3b_DFT_1989,3b_DFT_1990,Hirayama_GWDFT_2018}.

To take into account the inter-electronic correlations resulting from the significant onsite Coulomb repulsion at the copper atomic sites, we use the approach based on the so-called diagrammatic expansion of the Gutzwiller wave function (DE-GWF method). The method has been applied to both single- and multi-band models \cite{Zegrodnik_1band_1,Zegrodnik_1band_2,Kaczmarczyk_2013,Kaczmarczyk2014,MWysokinski_Anderson,2band_Hub_Bunemann} as well as recently to the description of the superconducting phase of the cuprates within the three band Emery model\cite{Zegrodnik_3band}, which is also considered here.

The Gutzwiller-type projected many particle wave function is taken in the form
\begin{equation}
 |\Psi_G\rangle\equiv\hat{P}|\Psi_0\rangle=\prod_{il}\hat{P}_{il}|\Psi_0\rangle \;,
 \label{eq:GWF}
\end{equation}
where $|\Psi_0\rangle$ represents the wave function of uncorrelated state. The intra-site intra-orbital projection operator has the following form
\begin{equation}
 \hat{P}_{il}\equiv \sum_{\Gamma}\lambda_{\Gamma|il}|\Gamma\rangle_{il\;il}\langle\Gamma|\;,
 \label{eq:P_Gamma}
\end{equation}
where $\lambda_{\Gamma|il}$ are the variational parameters determining relative weights corresponding to $|\Gamma\rangle_{il}$, which in turn represent states of the local basis on the atomic sites with the three types of orbitals ($l\in\{{d,p_x,p_y}\}$)
\begin{equation}
|\Gamma\rangle_{il}\in \{|\varnothing\rangle_{il}, |\uparrow\rangle_{il}, |\downarrow\rangle_{il},
|\uparrow\downarrow\rangle_{il}\}\;.
\label{eq:local_states}
\end{equation}
The consecutive states represent the empty, singly, and doubly occupied local configurations, respectively. As can be seen, the variational parameters, which tune the local electronic configurations in the resulting wave function, are orbital-dependent. By minimizing the energy of the system over the variational parameters one reduces the number of configurations which correspond to increased interaction energies. The details of the DE-GWF calculation scheme as applied to the $d$-$wave$ superconducting phase in the three-band Emery model are provided in Ref. \onlinecite{Zegrodnik_3band}. It should be noted, that the $C_4$ symmetry breaking leads to much more involved calculations, since the number of the so-called hopping and pairing lines which determine the $|\Psi_0\rangle$ wave function (cf. Ref. \onlinecite{Zegrodnik_3band,Zegrodnik_1band_1} for the definition of the lines) is twice as large as that corresponding to the situation in which the $C_4$ symmetry is preserved.

In the considered model the nematicity is realized by the $p$-orbital polarization which means that $n_{p_x}\neq n_{p_y}$ within each unit cell. Thus, the $i$ site index in the variational parameters $\lambda_{\Gamma|il}$ can be dropped and we end up with three sets of variational parameters $\lambda_{\Gamma|d}$, $\lambda_{\Gamma|p_x}$, and $\lambda_{\Gamma|p_y}$, which correspond to different electronic configurations at the three orbitals appearing in the model. As we have shown in Ref. \onlinecite{Zegrodnik_3band}, due to the fact that the Coulomb repulsion at the copper orbitals is the dominant energy in the system, the projection at the oxygen orbitals can be omitted by taking $\lambda_{\Gamma|p_x}=\lambda_{\Gamma|p_y}\equiv 1$. This assumption is also applied here. However, since the oxygen degrees of freedom are particularly important for the creation of the nematic phase, in the Appendix A we show explicitly that the results are not altered significantly by including the $p$-orbital projection

\section{Results and discussion}
\subsection{Intra-unit-cell nematicity}
In this subsection, we analyze the spontaneous formation of the intra-unit-cell nematicity without the inclusion of superconducting pairing. The effect of the latter is studied in the next subsection. In all the figures hole doping is defined in the following manner: $\delta = 5-n_{p_x}-n_{p_y}-n_d$, hence the parent compound corresponds to five electrons on each CuO$_2$ complex. In the nematic phase, the electronic concentration is shifted between the $p_x$ and $p_y$ orbitals, which induces the $C_4$ symmetry breaking. The corresponding nematic order parameter is thus defined in the following manner: 
\begin{equation}
\eta \equiv(n_{p_x}-n_{p_y})/(n_{p_x}+n_{p_y}),
\label{eq:eta}
\end{equation}
and represents the normalized $p$-orbital polarization. For nonzero $\eta$ also the $d$-$p$ hopping expectation values in the $(1,0)$ and $(0,1)$ directions differ. The parameter corresponding to the latter effect is defined in the analogous manner
\begin{equation}
\gamma = (P_{dp_x}-P_{dp_y})/(P_{dp_x}+P_{dp_y}),
\label{eq:gamma}
\end{equation}
where $P_{dp_x}$ and $P_{dp_y}$ are the nearest-neighbor hopping expectation values in the correlated state $|\Psi_G\rangle$. They are defined in the following manner: $P_{dp_x}=\langle\hat{c}^{\dagger}_{id\sigma}\hat{c}_{ip_x\sigma} \rangle_G$ and $P_{dp_y}=\langle\hat{c}^{\dagger}_{id\sigma}\hat{c}_{ip_y\sigma} \rangle_G$, where $\langle...\rangle_G=\langle\Psi_G|...|\Psi_G\rangle/\langle\Psi_G|\Psi_G\rangle$. We carry out our analysis for the typical values of the model parameters. If not stated otherwise they are set to: $t_{dp}=1.13\;$eV, $t_{pp}=0.49\;$eV, $\epsilon_{dp}=\epsilon_d-\epsilon_p=3.2\;$eV, $U_d=8\;$ eV, $U_p=4.1\;$eV.

\begin{figure}
 \centering
 \includegraphics[width=0.50\textwidth]{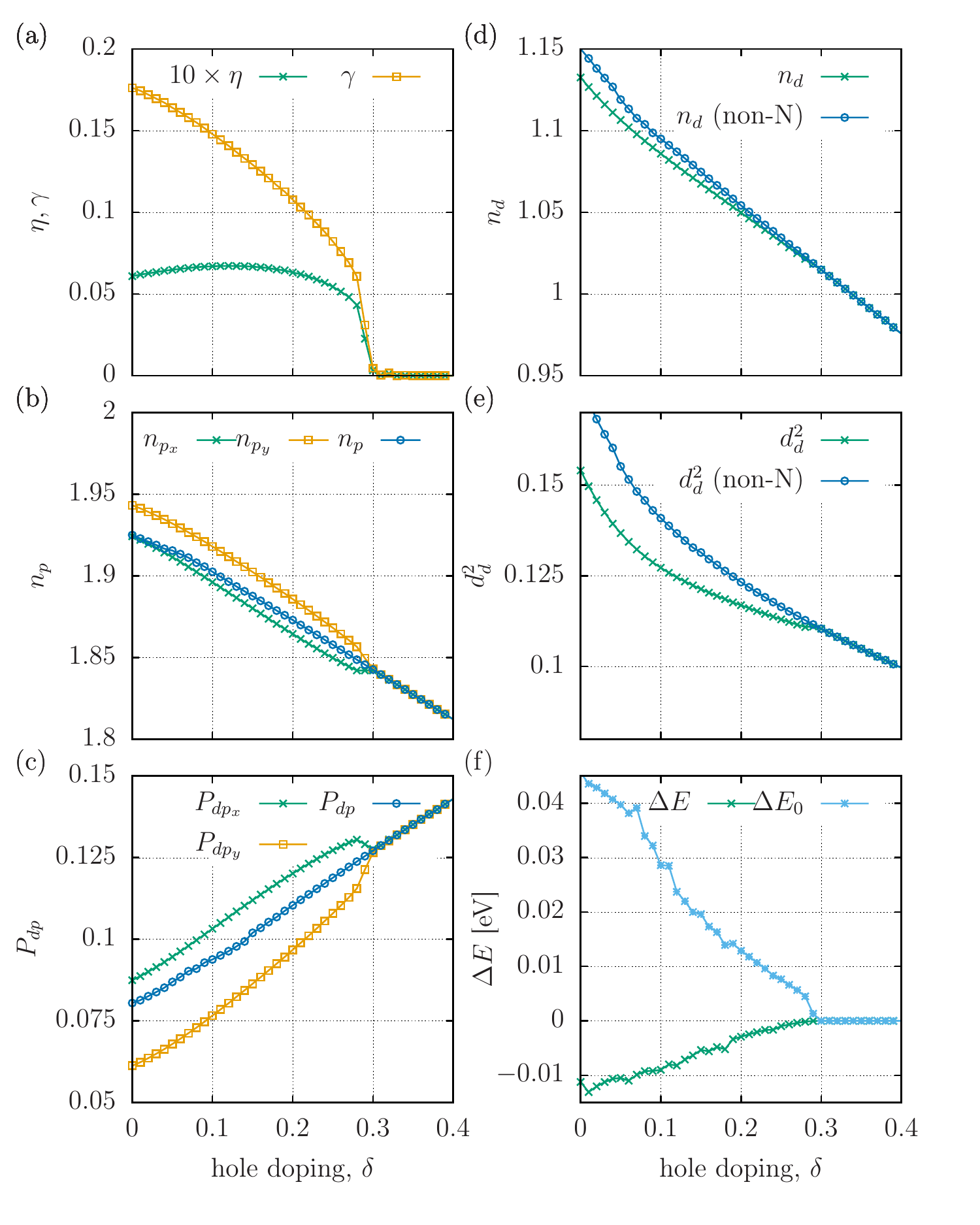}
 \caption{(a) The nematic order parameter $\eta$ together with the hopping asymmetry parameter $\gamma$, defined by Eqs. (\ref{eq:eta}) and (\ref{eq:gamma}) respectively, as a function of hole doping; (b-e) The  electronic concentrations $n_{p_x}$, $n_{p_y}$, $n_{d}$, the hopping expectation values $P_{dp_x}$, $P_{dp_x}$, and the double occupancies at the $d$-orbital $d^2_d$, all as functions of hole doping. Additionally, the values of all the physical quantities in (b-e) for the case of non-nematic (normal) phase are also provided (in blue). They are: $n_{p_x}=n_{p_y}\equiv n_{p}$, $P_{dp_x}=P_{dp_y}\equiv P_{dp}$, $n_d$ (non-N), $d^2_d$ (non-N); (f) The energy difference between the nematic phase and the non-nematic phase $\Delta E=E_N-E_{\mathrm{non-N}}$ as well as the corresponding kinetic energy contribution $\Delta E_0$, both vs. hole doping. }
 \label{fig:ndep_Nem}
\end{figure}

As we show in Fig. \ref{fig:ndep_Nem} the nematic phase appears in a significant hole doping range below $\delta\lesssim 0.3$, where both $\eta\neq 0$ and $\gamma\neq 0$. Relatively small normalized $p$-orbital polarization ($\eta$) induces significantly larger values of the normalized hopping asymmetry ($\gamma$). This can be also seen in (b) and (c) where we show explicitly the values of the electronic concentrations $n_{p_x}$, $n_{p_y}$ and hopping expectation values $P_{dp_x}$,$P_{dp_x}$. For the sake of comparison, in (b-e) we provide the corresponding results for the non-nematic state with the $C_4$ symmetry constraint ($n_{p_x}=n_{p_y}\equiv n_{p}$, $P_{dp_x}=P_{dp_y}\equiv P_{dp}$). As seen in (d), apart from the usual $p$-orbital polarization at the transition to the nematic state, there is also a relatively small electron concentration transfer from $d$ to $p$ orbital. This results in a reduced number of double occupancies at the $d$ orbitals ($d^2_{d}\equiv \langle\hat{n}_{id\uparrow}\hat{n}_{id\downarrow} \rangle_G$) in the nematic state with respect to the normal, non-nematic state (e). The latter effect decreases the interaction energy resulting from the Coulomb repulsion at the copper atomic sites. However, the interaction energy loss at the transition to the nematic phase is at the expense of the kinetic energy gain. Nevertheless, the overall energy balance is advantageous leading to the nematic behavior of the system. This is explicitly shown in (f) where the energy difference between the nematic and non-nematic states is plotted ($\Delta E=E_N-E_{\mathrm{non-N}}$) as a function of hole doping. Additionally, the kinetic energy gain is also provided in the Figure ($\Delta E_0$). 

One should note that the only interaction terms included in the model have an onsite character, with the dominant contribution coming from the Coulomb repulsion on the copper atomic sites. Therefore, the appearance of nematicity shown in Fig. \ref{fig:ndep_Nem} suggests that the electron correlations resulting from the high $U_d$ value play the dominant role in the spontaneous $C_4$ symmetry breaking. Such a conclusion is distinct from the analysis presented in Refs. \onlinecite{Kivelson2004,Fischer2011,Bulut2013,Tsuchiizu2018}, where the role of the intersite oxygen-oxygen Coulomb repulsion in creating the nematic phase has been emphasized.

\begin{figure}
 \centering
 \includegraphics[width=0.50\textwidth]{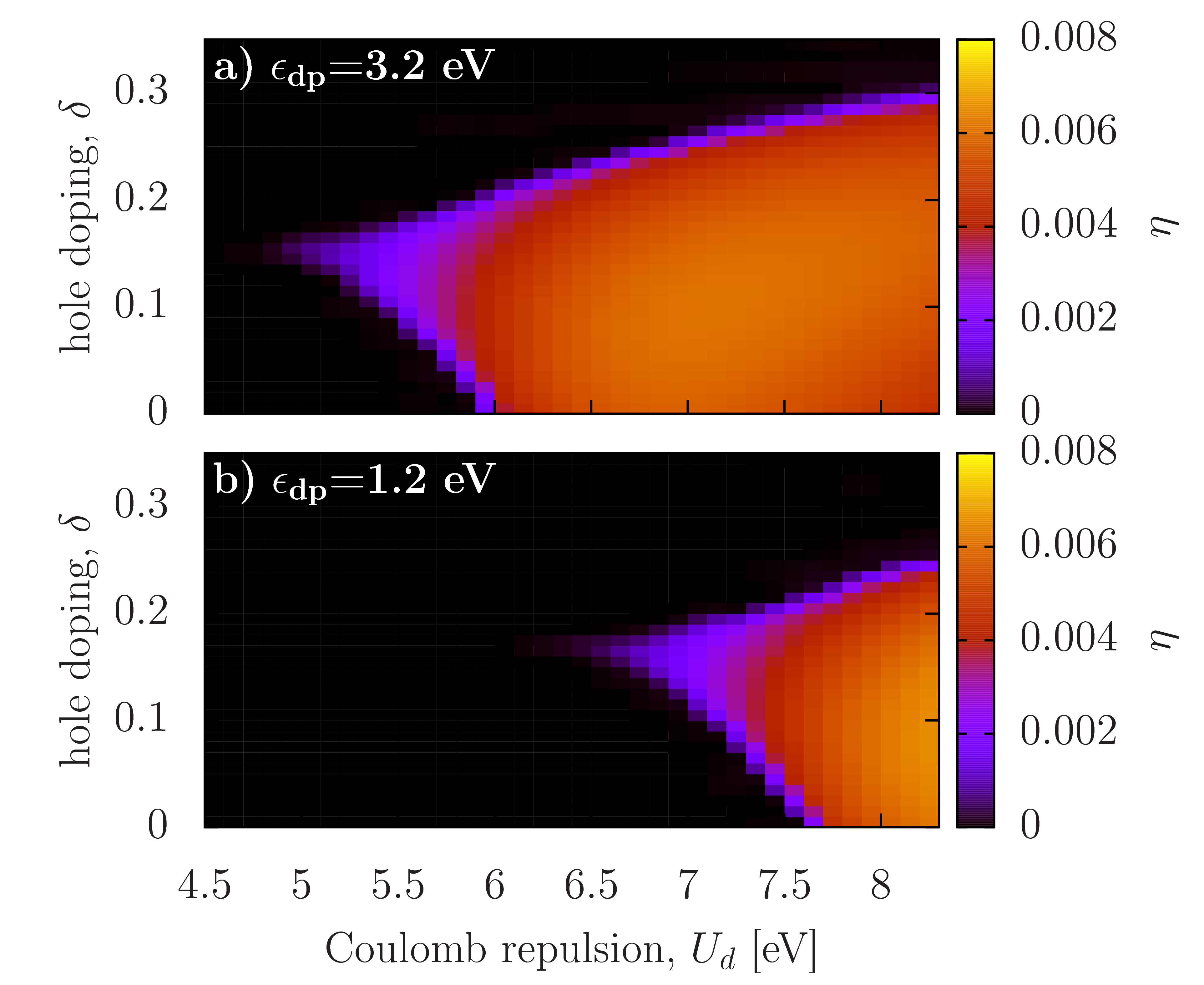}
 \caption{The nematic order parameter [cf. Eq. (\ref{eq:eta})] as a function of both $U_d$ and $\delta$ for two selected values of $\epsilon_{dp}=3.2$ and $1.2\;$eV.}
 \label{fig:nU_diag}
\end{figure}

\begin{figure}
 \centering
 \includegraphics[width=0.49\textwidth]{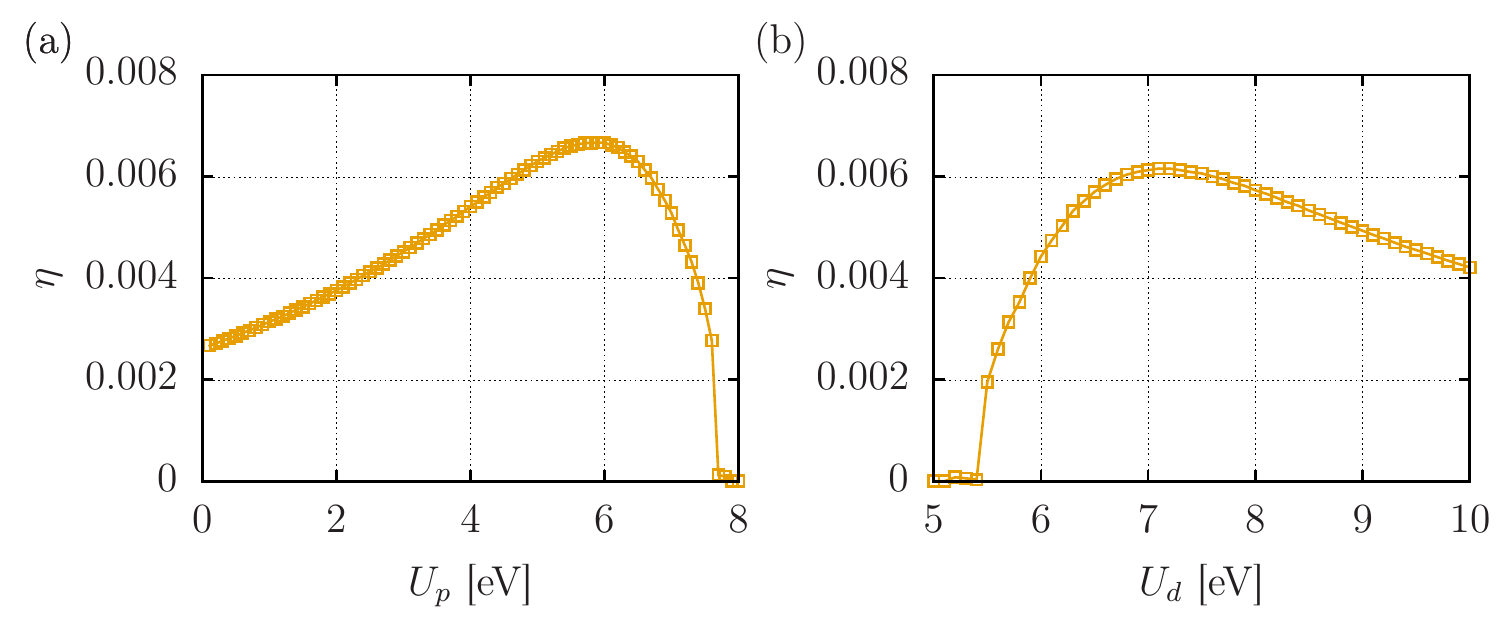}
 \caption{(a) Nematic order parameter as a function of $U_p$ for doping $\delta=0.1$ and $U_d=8.3\;$eV; (b) The same as a function of $U_d$ for doping $\delta=0.1$ and $U_p=4.1\;$eV.}
 \label{fig:Up_Ud_dep}
\end{figure}

To analyze in detail the influence of the $U_d$-term on the nematic behavior, in Fig. \ref{fig:nU_diag} we have plotted the order parameter $\eta$ on the $(U_d,\delta)$ plane. As one can see, the intrasite Coulomb repulsion integral has to be large enough to induce the onset of nematic phase, what again indicates the significant role of the onsite electronic correlations in the $C_4$ symmetry breaking. Furthermore, the region of the nematic phase stability is very similar to that corresponding to the superconducting phase stability determined by us very recently in Ref. \onlinecite{Zegrodnik_3band} (Fig. 12 in that paper) within the same model. Furthermore, by reducing the energy difference between the copper and oxygen atomic levels ($\epsilon_{dp}=\epsilon_d-\epsilon_p$) one moves the nematic phase stability regime towards larger values of $U_d$ [cf. Figs. \ref{fig:nU_diag} (a) and (b)]. Again, the same effect is seen for the case of the paired phase\cite{Zegrodnik_3band}. As discussed previously\cite{Zegrodnik_3band}, the lowest-energy excitation for the parent compound ($\Delta E=U_d-U_p+\epsilon_{dp}$) should be considered as the factor that determines the strength of the electronic correlations in the model. Therefore, by decreasing $\epsilon_{dp}$ one has to provide higher values of $U_d$ so that to achieve large enough $\Delta E$ to induce the nematicity. The similarity between nematic phase and superconducting phase behaviors in this respect points to the common origin of both. In the considered scenario such an origin would be the inter-electronic correlations, with the dominant contribution comming from the onsite Coulomb repulsion at the copper sites. This interpretation is also consistent with the determined $U_p$ dependence of the nematic order parameter (cf. Fig. \ref{fig:Up_Ud_dep}). The $U_d$ and $U_p$ parameters enter the expression for $\Delta E$ with opposite signs, what leads to the opposite effect of the two parameters on the order parameter $\eta$ seen in Figs. \ref{fig:Up_Ud_dep}. By increasing $U_p$ we decrease $\Delta E$, hence, for high enough $U_p$ values the correlation strength governed by $\Delta E$ is too small to induce nematicity and hence, $\eta$ reduces to zero.

\subsection{Coexistence of superconductivity and nematicity}
Since both the nematic ordering and the $d$-$wave$ superconductivity seem to have the same origin in the considered approach, and they reside at the similar area of the ($U_d$,$\delta$)-phase diagram (cf. Fig. \ref{fig:nU_diag} here and Fig. 12 in Ref. \onlinecite{Zegrodnik_3band}), the question of interplay between the two is in place here. Therefore, we have carried out calculations in which both the superconducting pairing and the $C_4$ symmetry breaking may appear together. As shown in our recent paper\cite{Zegrodnik_3band}, within the three-band description various pairing amplitudes contribute to the resultant superconducting phase. They correspond to the intra- and inter-orbital pairing between subsequent nearest-neighboring atomic sites. Nevertheless, the dominant contribution results from the pairing between the nearest-neighbor $d$-orbitals due to copper. Therefore, here we focus on the analysis of the latter pairing amplitude and do not show the remaining ones, which are much smaller. Since in the nematic phase the $(1,0)$ and $(0,1)$ directions within the Cu-O plane are not equivalent, a mixed $d$- and $s$-$wave$ pairing may appear in the coexistent superconducting-nematic phase. The correlated $d$-$wave$ and $s$-$wave$ gap parameters that are going to be analyzed have the following form
\begin{equation}
\begin{split}
    \Delta^{(d)}_{dd|i}&\equiv\frac{1}{4}\sum_j(-1)^{\rho^{d}_{ij}}\Delta^{(i,j)}_{dd}\\
    \Delta^{(s)}_{dd|i}&\equiv\frac{1}{4}\sum_j\Delta^{(i,j)}_{dd},
\end{split}
\label{eq:deltas_s_d}
\end{equation}
where the summations run over the nearest-neighbor $d$-$d$ orbitals only, $\Delta^{(i,j)}_{dd}=\langle\hat{c}^{\dagger}_{id\uparrow}\hat{c}^{\dagger}_{jd\downarrow} \rangle_G$ and
\begin{equation}
  \rho^{d}_{ij}=\begin{cases}
    0, & \text{if } \vec{R}_{ij}=(1,0) \text{ or } \vec{R}_{ij}=(-1,0),\\
    1, & \text{if } \vec{R}_{ij}=(0,1) \text{ or } \vec{R}_{ij}=(0,-1),
  \end{cases}
\end{equation}
with $\vec{R}_{ij}=\vec{R}_i-\vec{R}_j$. Since we are considering a homogeneous situation, the $i$ indices in the expressions for the gap parameters can be dropped ($\Delta^{(d)}_{dd|i}\equiv\Delta^{(d)}_{dd}$, $\Delta^{(s)}_{dd|i}\equiv\Delta^{(s)}_{dd}$).

\begin{figure}
 \centering
 \includegraphics[width=0.49\textwidth]{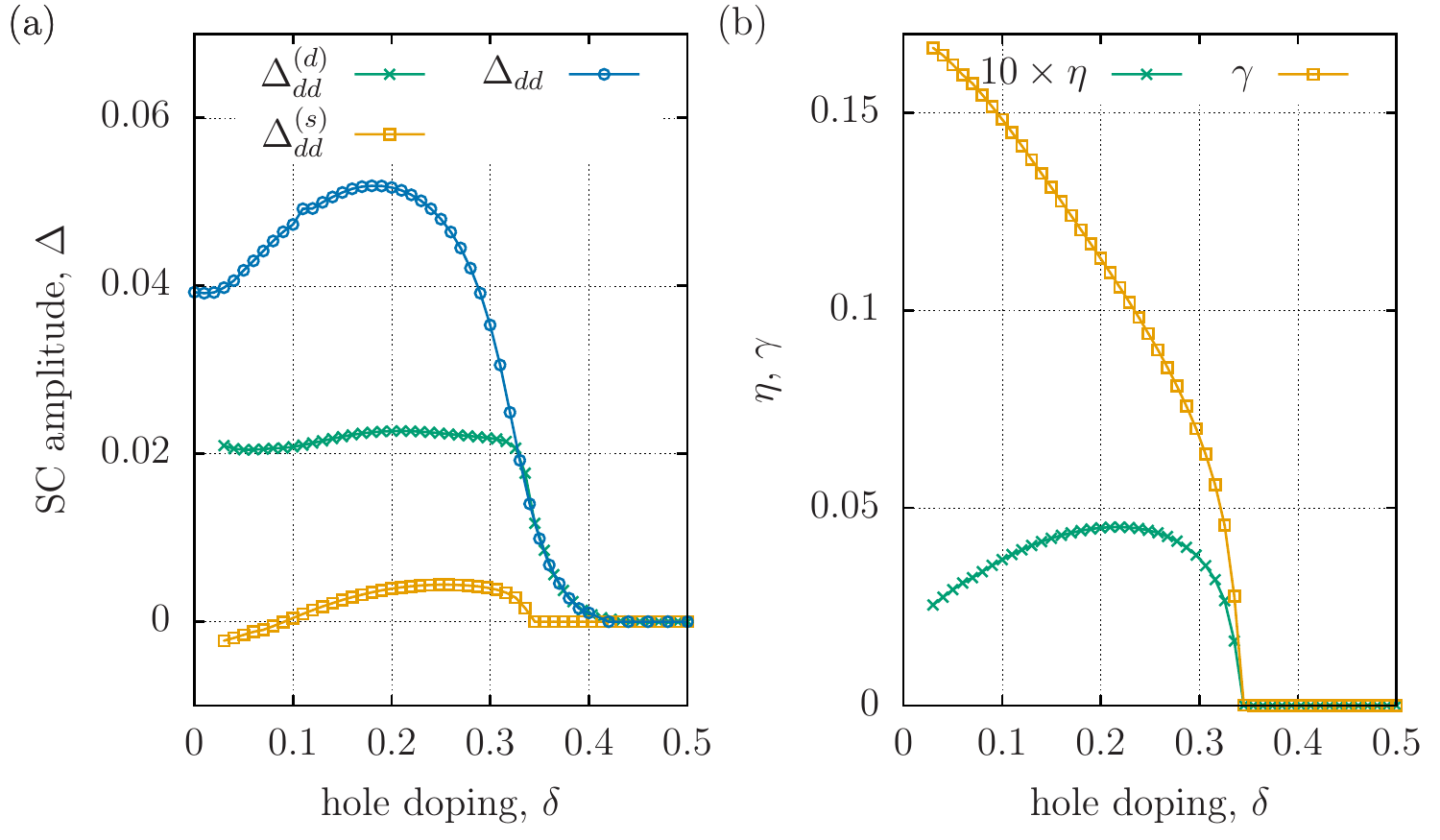}
 \caption{(a) $d$- and $s$-wave pairing amplitudes between the $d$-$d$ nearest neighbor atomic sites as a function of hole doping for the case of the coexistent superconducting-nematic phase. Additionally, in blue the $d$-$d$ pairing amplitude is shown for the case of the superconducting phase without the nematicity where the pairing is purely of the $d$-$wave$ character; (b) The nematic order parameter $\eta$ tohether with $\gamma$ (cf. Eqs. \ref{eq:eta} and \ref{eq:gamma}) also as functions of hole doping. The results have been obtained for $U_d=10.3\;$eV, $\epsilon_{dp}=3.2\;$ eV, and $U_p=4.1\;$eV.}
 \label{fig:delt_nematic}
\end{figure}

\begin{figure}
 \centering
 \includegraphics[width=0.49\textwidth]{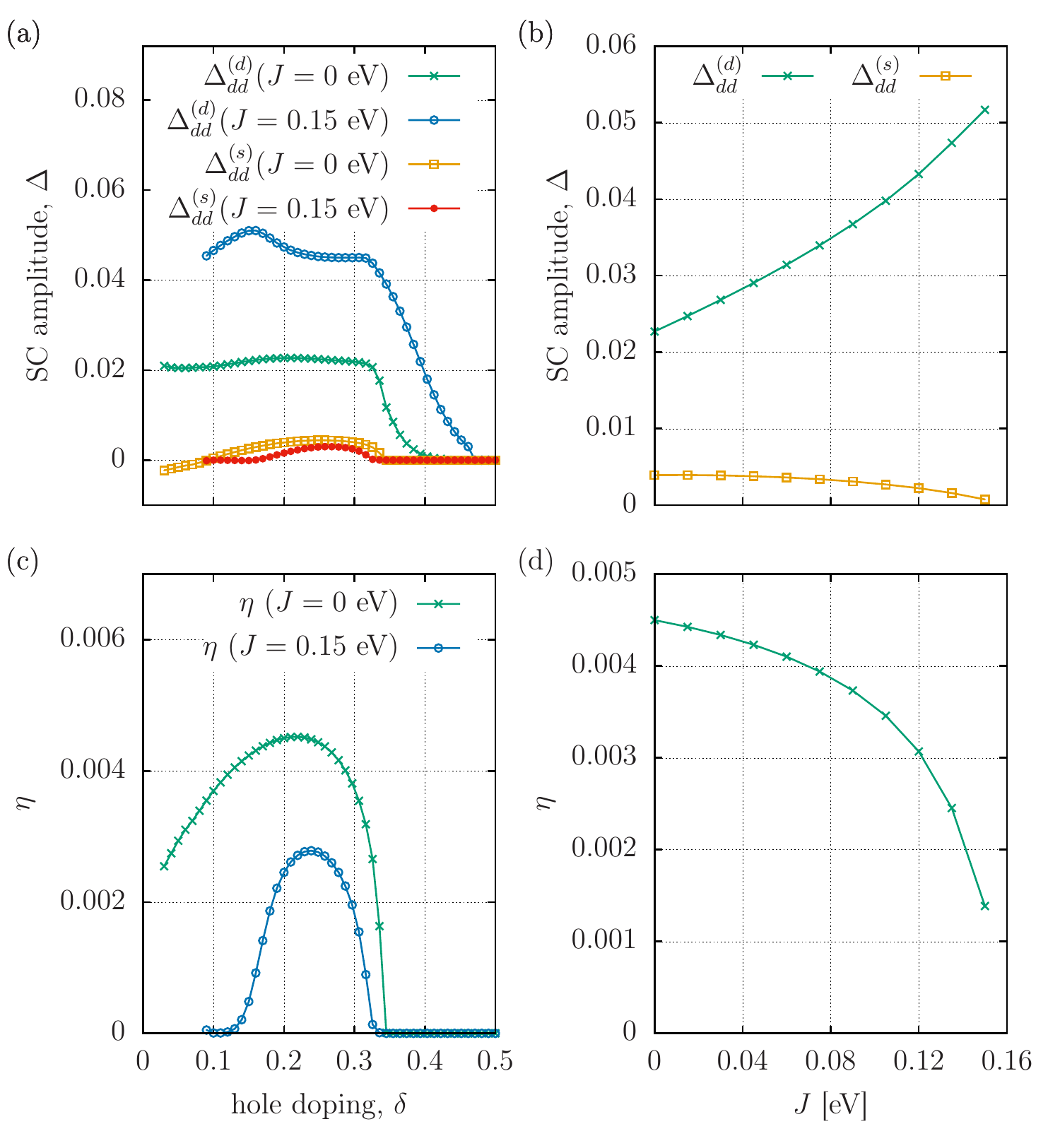}
 \caption{(a) $d$- and $s$-wave pairing amplitudes between $d$-$d$ nearest-neighbor atomic sites as a function of hole doping for two selected values of the exchange integral $J=0.0\;$eV and $J=0.15\;$ eV; (b) The same as in (a) but as a function of the exchange integral for $\delta=0.2$; (c) The nematic order parameter $\eta$ as a function of hole doping for two selected values of $J$; (d) The same as in (c) but as a function of the exchange integral for $\delta=0.2$. For meaning of the exchange integral see the main text.}
 \label{fig:delt_J}
\end{figure}

As shown in Fig. \ref{fig:delt_nematic}, the $d$- and $s$-$wave$ pairing amplitudes, as well as the nematic order parameter $\eta$, all become non-zero in the doping range below $\sim0.3$, which indicates the appearance of the coexistent superconducting-nematic phase (SC+N). For comparison, in Fig. \ref{fig:delt_nematic} (a) we show the $d$-$d$ pairing amplitude for the case of pure superconducting $d$-$wave$ state for which the $C_4$ symmetry is preserved (in blue). Above $\delta\approx 0.3$ the $d$-$wave$ SC gap increases with decreasing doping, however, below $\delta\approx0.3$ where the nematicity sets in the $\Delta^{(d)}_{dd}$ becomes very weakly dependent on the doping and is significantly reduced with respect to the gap corresponding to the pure SC state (cf. green and blue lines in Fig. \ref{fig:delt_nematic}a). Additionally, in the region of the SC+N phase stability, a small $s$-$wave$ contribution to the pairing appears (yellow line in Fig. 
\ref{fig:delt_nematic}a). It can be concluded from the experimental research that superconductivity and nematicity appear simultaneously in a wide doping range reaching above the optimal doping for $T\approx 0$ K in the cuprates\cite{Lawler2010,Hashimoto2014,Sato2017}. However, it is not clear if in fact the coexistent superconducting-nematic phase appears in the experiments or a phase separation scenario is realized with a purely nematic non-superconductig domains residing inside an essentially $d$-$wave$ superconducting environment. 

As we have show, within the present approach the two broken-symmetry states can coexist (cf. Fig. \ref{fig:delt_nematic}). Nevertheless, the suppression of the $d$-$wave$ pairing amplitude in the SC+N phase should be considered as a signature of competition between the $d$-$wave$ superconductivity and nematic order, which is also seen in the single-band models of the Cu-O plane\cite{Zegrodnik2018}. In the latter model, the exchange term between the nearest-neighbor atomic sites works in favor of the superconducting phase, reducing the nematic behavior. In order to analyze if a similar effect can be seen here we have carried out the calculations for the SC+N phase in the three-band model (\ref{eq:Hamiltonian_start}) with the inclusion of a similar exchange term between the copper atomic-sites
\begin{equation}
    \hat{H}_J=J\sum_{\langle ij\rangle}\hat{\mathbf{S}}_{id}\cdot\hat{\mathbf{S}}_{jd},
    \label{eq:J_term}
\end{equation}
where $J$ is the exchange integral, the summation is carried out over the nearest-neighbor copper atomic sites, and $\hat{\mathbf{S}}_{id}$ are the spin operators at the $d$ orbitals. The estimates for the $J$ value via Raman scattering experiments for the undoped situation varies between $0.1\;$eV and $0.14\;$ eV, depending on particular compound\cite{Lyons1988,Sugai1988,Blumberg1996}, which is consistent with the theoretical estmates\cite{Mizuno1998,tJmodel_rev_2008}. The model defined by Eq. (\ref{eq:Hamiltonian_start}) supplemented with the term given by (\ref{eq:J_term}) constitutes the three-band correspondent of the so-called $t$-$J$-$U$ model\cite{Zegrodnik_1band_2}. 

As one can see in Fig. \ref{fig:delt_J} for non-zero values of $J$ the $d$-$wave$ pairing amplitude is enhanced in the wide range of the hole doping, in contrast to both the $s$-$wave$ pairing amplitude and the nematic order parameter. Above $J\approx 0.16\;$eV the latter is completely suppressed and the stability of the pure SC phase is restored. Nevertheless, as shown previously\cite{Fischer2011} the inter-site oxygen-oxygen Coulomb repulsion $\sim V$ strengthens the nematic phase. Therefore, one can expect that the $V$-term can lead to the appearance of the nematic behavior even for $J\approx 0.16\;$eV. Hence, the final form of the ground state with respect to the $C_4$ symmetry breaking may result from a subtle interplay between the two factors.

\section{Conclusions}

We have shown that the nematic phase can appear in the three-band Emery model in the absence of the inter-site oxygen-oxygen Coulomb repulsion, which has been considered as the triggering effect of nematicity in the previous study\cite{Kivelson2004,Fischer2011,Bulut2013}. Also, as shown here, at the transition to the nematic phase electron concentration transfer between the $d$- and $p$- orbitals takes place in addition to the usually discussed $p_x/p_y$ polarization (cf. Fig. \ref{fig:ndep_Nem}e). Such an effect leads to a decrease of the number of double occupancies on the $d$ orbitals, which is energetically favorable due to the strong onsite Coulomb repulsion at those orbitals. 

According to our analysis, a spontaneous $C_4$ symmetry breaking appears due to inter-electronic correlations, strength of which is determined by the energy value corresponding to the electron transfer from the oxygen $p$- to the copper $d$-orbitals (for the parent compound $\Delta E=U_d-U_p+\epsilon_{dp}$). A significant value of $\Delta E$ has to be reached to induce the nematic phase - a condition that is met for the model parameters corresponding to the copper-oxides. Such an interpretation is consistent with the fact that by decreasing $\epsilon_{dp}$ one moves the whole nematic phase stability regime towards higher $U_d$ values (cf. Fig. \ref{fig:nU_diag}). Also, the effect of $U_d$ and $U_p$ parameters on the nematic phase is of opposite character, due to opposite signs of the two factors when entering the $\Delta E$ expression (cf. Fig. \ref{fig:Up_Ud_dep}).

The results analyzed here and those presented in our previous report\cite{Zegrodnik_3band} point to a common origin of both the superconducting and nematic phase (cf. Fig. \ref{fig:nU_diag} here and Fig. 12 in Ref. \onlinecite{Zegrodnik_3band}). Also, we show that the superconducting and nematic phases may coexist in a significant doping range leading to a suppression of the $d$-$wave$ pairing amplitude and the appearance of a small $s$-$wave$ contribution to the pairing. Similarly as in the single band picture, the competition between the $d$-$wave$ pairing and $C_4$ symmetry breaking may be tuned by the exchange interaction term and the intersite Coulomb repulsion with the former working in favor of the pairing and the later enhancing nematicity. 

One should note that according to experimental research both superconductivity and the pseudogap phase appear in a wide doping range (cf. Fig 6d in Ref. \onlinecite{Hashimoto2014}) reaching above the optimal doping for $T\ll T_C$ (with $T_C$ being the superconducting critical temperature). On the other hand, a strong evidence of nematicity in the pseudogap state has been provided quite recently\cite{Sato2017}. Such experimental picture could be reconciled with the results presented here, where the superconducting-nematic coexistent phase appears also in a relatively wide doping range for $T=0$ K (cf. Fig. \ref{fig:delt_nematic}). Furthermore, the weak doping dependence of the correlated $d$-$wave$ pairing amplitude inside the SC+N phase shown in Fig. \ref{fig:delt_nematic}a can be related to the experimental result presented in Fig 6e of Ref. \onlinecite{Hashimoto2014}, where it is reported that the gap slope near the nodal direction, corresponding to the $d$-$wave$ symmetry of the pairing, shows a similar behavior due to the presence of the pseudogap phase. On the other hand, the scenario with no coexistence region but with purely nematic domains residing inside a $d$-$wave$ superconducting environment, would be in accord with the measured $d$-$wave$ symmetry of the gap in the whole doping range where the pairing appears. Moreover, it should be noted that within our approach we do not analyze directly the pseudogap behavior. Therefore, the definite answer to the question of the relation between the nematic phase and the pseudogap behaviour is beyond the scope of this paper.

At the end, one should note that our results correspond to the ground state of the system ($T=0$ K). A natural question concerns the effect of finite temperatures on both the pairing and $C_4$ symmetry breaking considered by us here. Unfortunately, the Gutzwiller wave function is designed to describe the ground state properties and the application of such an approach to non-zero temperatures is not resolved as yet.

% Answer to this question would allow for validation of the discussed connection between superconductivity, nematicity and pseudogap state by comparing the critical temperatures for the formation of the analyzed phases with the experimental ones.

%One should note that by projecting out the double occupacies at the $d$-orbitals one can derive other exchange-type terms which also may have influence on the appearance of the nematic phase\cite{tJmodel_rev_2008} but their inclusion is beyond the scope of this analysis. 

%In the three-band picture such term appear as a result of projecting out the double occupancies via canonical transformation which results in $J=0.12\;$eV, which is consistant with the experimental value of $J=0.13\;$eV.

\section{Acknowledgement}
M.Z. and A.B. acknowledge the financial support through the Grant SONATA, No. 2016/21/D/ST3/00979 from the National Science Centre (NCN), Poland. J.S.  acknowledges  the  financial  support  by the Grant OPUS No. UMO-2018/29/B/ST3/02646 from the National Science Centre (NCN), Poland.

\appendix
\section{Effect of the $p$-orbital Coulomb repulsion on nematicity}
\begin{figure}
 \centering
 \includegraphics[width=0.49\textwidth]{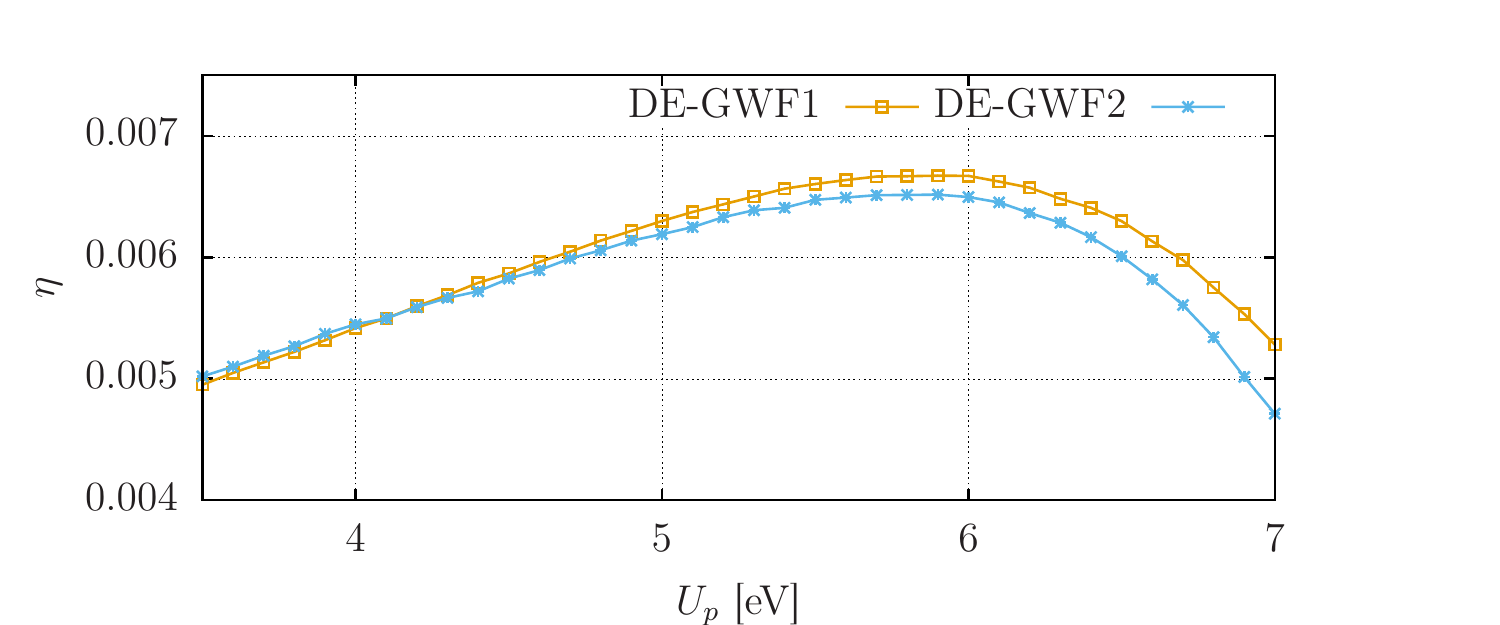}
 \caption{Nematic order parameter as a function of $U_p$ for doping $\delta=0.1$ and $U_d=8.3\;$eV obtained by using the two different calculation schemes, DE-GWF1 and DE-GWF2. For the former the Gutzwiller-type projection is carried out only at the copper atomic sites, while for the latter the full projection on both copper and oxygen atomic sites is applied.}
 \label{fig:xp_comparison}
\end{figure}

Here, we show that within the presented approach one can safely neglect the projection at the oxygen atomic sites when considering the nematic phase in the three-band Emery model. In Fig. \ref{fig:xp_comparison} we show the nematic order parameter as a function of the $p$-orbital Coulomb repulsion for the case of the two calculation schemes, DE-GWF1 and DE-GWF2. For the former the projection at the $p$-orbitals has been neglected, while for the latter it was included. For DE-GWF1 we set $\lambda_{\Gamma|px}=\lambda_{\Gamma|py}\equiv 1$ [cf. Eq. (\ref{eq:P_Gamma})], while for DE-GWF2, $\lambda_{\Gamma|px}$ and $\lambda_{\Gamma|py}$ are treated as variational parameters over which the energy is minimized. As one can see the $U_p$ dependence of $\eta$ is similar in both cases and for the parameter range corresponding to the cuprates ($U_p\approx4-6\;$ eV) the obtained results are virtually the same. Similar results cqn be obtained for other hole dopings.

\bibliography{refs.bib}

%merlin.mbs apsrev4-1.bst 2010-07-25 4.21a (PWD, AO, DPC) hacked
%Control: key (0)
%Control: author (8) initials jnrlst
%Control: editor formatted (1) identically to author
%Control: production of article title (-1) disabled
%Control: page (0) single
%Control: year (1) truncated
%Control: production of eprint (0) enabled
\begin{thebibliography}{36}%
\makeatletter
\providecommand \@ifxundefined [1]{%
 \@ifx{#1\undefined}
}%
\providecommand \@ifnum [1]{%
 \ifnum #1\expandafter \@firstoftwo
 \else \expandafter \@secondoftwo
 \fi
}%
\providecommand \@ifx [1]{%
 \ifx #1\expandafter \@firstoftwo
 \else \expandafter \@secondoftwo
 \fi
}%
\providecommand \natexlab [1]{#1}%
\providecommand \enquote  [1]{``#1''}%
\providecommand \bibnamefont  [1]{#1}%
\providecommand \bibfnamefont [1]{#1}%
\providecommand \citenamefont [1]{#1}%
\providecommand \href@noop [0]{\@secondoftwo}%
\providecommand \href [0]{\begingroup \@sanitize@url \@href}%
\providecommand \@href[1]{\@@startlink{#1}\@@href}%
\providecommand \@@href[1]{\endgroup#1\@@endlink}%
\providecommand \@sanitize@url [0]{\catcode `\\12\catcode `\$12\catcode
  `\&12\catcode `\#12\catcode `\^12\catcode `\_12\catcode `\%12\relax}%
\providecommand \@@startlink[1]{}%
\providecommand \@@endlink[0]{}%
\providecommand \url  [0]{\begingroup\@sanitize@url \@url }%
\providecommand \@url [1]{\endgroup\@href {#1}{\urlprefix }}%
\providecommand \urlprefix  [0]{URL }%
\providecommand \Eprint [0]{\href }%
\providecommand \doibase [0]{http://dx.doi.org/}%
\providecommand \selectlanguage [0]{\@gobble}%
\providecommand \bibinfo  [0]{\@secondoftwo}%
\providecommand \bibfield  [0]{\@secondoftwo}%
\providecommand \translation [1]{[#1]}%
\providecommand \BibitemOpen [0]{}%
\providecommand \bibitemStop [0]{}%
\providecommand \bibitemNoStop [0]{.\EOS\space}%
\providecommand \EOS [0]{\spacefactor3000\relax}%
\providecommand \BibitemShut  [1]{\csname bibitem#1\endcsname}%
\let\auto@bib@innerbib\@empty
%</preamble>
\bibitem [{\citenamefont {M.~J.~Lawler}\ and\ \citenamefont
  {Kim}(2010)}]{Lawler2010}%
  \BibitemOpen
  \bibfield  {author} {\bibinfo {author} {\bibfnamefont {J.~L. A. R. S. Y. K.
  C. K. K. H. E. S. U. J. C. D. J. P.~S.}\ \bibnamefont {M.~J.~Lawler},
  \bibfnamefont {K.~Fujita}}\ and\ \bibinfo {author} {\bibfnamefont {E.-A.}\
  \bibnamefont {Kim}},\ }\href {\doibase 10.1038/nature09169} {\bibfield
  {journal} {\bibinfo  {journal} {Nature}\ }\textbf {\bibinfo {volume} {466}},\
  \bibinfo {pages} {347} (\bibinfo {year} {2010})}\BibitemShut {NoStop}%
\bibitem [{\citenamefont {Benjamin A.~Frandsen}(2014)}]{Frandsen2014}%
  \BibitemOpen
  \bibfield  {author} {\bibinfo {author} {\bibfnamefont {H.~H. Y. Z. Y. N. H.
  K. Y. J. U. W.-G. Y. . S. J. L.~B.}\ \bibnamefont {Benjamin A.~Frandsen},
  \bibfnamefont {Emil S.~Bozin}},\ }\href {\doibase doi.org/10.1038/ncomms6761}
  {\bibfield  {journal} {\bibinfo  {journal} {Nat. Commun.}\ }\textbf {\bibinfo
  {volume} {5}},\ \bibinfo {pages} {5761} (\bibinfo {year} {2014})}\BibitemShut
  {NoStop}%
\bibitem [{\citenamefont {Achkar}\ \emph {et~al.}(2016)\citenamefont {Achkar},
  \citenamefont {Zwiebler}, \citenamefont {McMahon}, \citenamefont {He},
  \citenamefont {Sutarto}, \citenamefont {Djianto}, \citenamefont {Hao},
  \citenamefont {Gingras}, \citenamefont {H{\"u}cker}, \citenamefont {Gu},
  \citenamefont {Revcolevschi}, \citenamefont {Zhang}, \citenamefont {Kim},
  \citenamefont {Geck},\ and\ \citenamefont {Hawthorn}}]{Achkar2016}%
  \BibitemOpen
  \bibfield  {author} {\bibinfo {author} {\bibfnamefont {A.~J.}\ \bibnamefont
  {Achkar}}, \bibinfo {author} {\bibfnamefont {M.}~\bibnamefont {Zwiebler}},
  \bibinfo {author} {\bibfnamefont {C.}~\bibnamefont {McMahon}}, \bibinfo
  {author} {\bibfnamefont {F.}~\bibnamefont {He}}, \bibinfo {author}
  {\bibfnamefont {R.}~\bibnamefont {Sutarto}}, \bibinfo {author} {\bibfnamefont
  {I.}~\bibnamefont {Djianto}}, \bibinfo {author} {\bibfnamefont
  {Z.}~\bibnamefont {Hao}}, \bibinfo {author} {\bibfnamefont {M.~J.~P.}\
  \bibnamefont {Gingras}}, \bibinfo {author} {\bibfnamefont {M.}~\bibnamefont
  {H{\"u}cker}}, \bibinfo {author} {\bibfnamefont {G.~D.}\ \bibnamefont {Gu}},
  \bibinfo {author} {\bibfnamefont {A.}~\bibnamefont {Revcolevschi}}, \bibinfo
  {author} {\bibfnamefont {H.}~\bibnamefont {Zhang}}, \bibinfo {author}
  {\bibfnamefont {Y.-J.}\ \bibnamefont {Kim}}, \bibinfo {author} {\bibfnamefont
  {J.}~\bibnamefont {Geck}}, \ and\ \bibinfo {author} {\bibfnamefont {D.~G.}\
  \bibnamefont {Hawthorn}},\ }\href {\doibase 10.1126/science.aad1824}
  {\bibfield  {journal} {\bibinfo  {journal} {Science}\ }\textbf {\bibinfo
  {volume} {351}},\ \bibinfo {pages} {576} (\bibinfo {year} {2016})},\ \Eprint
  {http://arxiv.org/abs/https://science.sciencemag.org/content/351/6273/576.full.pdf}
  {https://science.sciencemag.org/content/351/6273/576.full.pdf} \BibitemShut
  {NoStop}%
\bibitem [{\citenamefont {Y.~Sato}(2017)}]{Sato2017}%
  \BibitemOpen
  \bibfield  {author} {\bibinfo {author} {\bibfnamefont {H.~M. Y. K. E.-G. M.
  T. N. T. L. J. P. B. K. T. S. . Y.~M.}\ \bibnamefont {Y.~Sato}, \bibfnamefont
  {S.~Kasahara}},\ }\href {\doibase 10.1038/nphys4205} {\bibfield  {journal}
  {\bibinfo  {journal} {Nature Physics}\ }\textbf {\bibinfo {volume} {13}},\
  \bibinfo {pages} {1074} (\bibinfo {year} {2017})}\BibitemShut {NoStop}%
\bibitem [{\citenamefont {Mesaros}\ \emph {et~al.}(2011)\citenamefont
  {Mesaros}, \citenamefont {Fujita}, \citenamefont {Eisaki}, \citenamefont
  {Uchida}, \citenamefont {Davis}, \citenamefont {Sachdev}, \citenamefont
  {Zaanen}, \citenamefont {Lawler},\ and\ \citenamefont {Kim}}]{Mesaros2011}%
  \BibitemOpen
  \bibfield  {author} {\bibinfo {author} {\bibfnamefont {A.}~\bibnamefont
  {Mesaros}}, \bibinfo {author} {\bibfnamefont {K.}~\bibnamefont {Fujita}},
  \bibinfo {author} {\bibfnamefont {H.}~\bibnamefont {Eisaki}}, \bibinfo
  {author} {\bibfnamefont {S.}~\bibnamefont {Uchida}}, \bibinfo {author}
  {\bibfnamefont {J.~C.}\ \bibnamefont {Davis}}, \bibinfo {author}
  {\bibfnamefont {S.}~\bibnamefont {Sachdev}}, \bibinfo {author} {\bibfnamefont
  {J.}~\bibnamefont {Zaanen}}, \bibinfo {author} {\bibfnamefont {M.~J.}\
  \bibnamefont {Lawler}}, \ and\ \bibinfo {author} {\bibfnamefont {E.-A.}\
  \bibnamefont {Kim}},\ }\href {\doibase 10.1126/science.1201082} {\bibfield
  {journal} {\bibinfo  {journal} {Science}\ }\textbf {\bibinfo {volume}
  {333}},\ \bibinfo {pages} {426} (\bibinfo {year} {2011})}\BibitemShut
  {NoStop}%
\bibitem [{\citenamefont {R.~Comin}(2015)}]{Comin2015}%
  \BibitemOpen
  \bibfield  {author} {\bibinfo {author} {\bibfnamefont {E.~H. d. S. N.-L. C.
  R. L. W. N. H. D. A. B. F. H. G. A. S. A.~D.}\ \bibnamefont {R.~Comin},
  \bibfnamefont {R.~Sutarto}},\ }\href {\doibase doi.org/10.1038/nmat4295}
  {\bibfield  {journal} {\bibinfo  {journal} {Science}\ }\textbf {\bibinfo
  {volume} {347}},\ \bibinfo {pages} {1335} (\bibinfo {year}
  {2015})}\BibitemShut {NoStop}%
\bibitem [{\citenamefont {Cyr-Choini\`ere}\ \emph {et~al.}(2015)\citenamefont
  {Cyr-Choini\`ere}, \citenamefont {Grissonnanche}, \citenamefont {Badoux},
  \citenamefont {Day}, \citenamefont {Bonn}, \citenamefont {Hardy},
  \citenamefont {Liang}, \citenamefont {Doiron-Leyraud},\ and\ \citenamefont
  {Taillefer}}]{Cyr2015}%
  \BibitemOpen
  \bibfield  {author} {\bibinfo {author} {\bibfnamefont {O.}~\bibnamefont
  {Cyr-Choini\`ere}}, \bibinfo {author} {\bibfnamefont {G.}~\bibnamefont
  {Grissonnanche}}, \bibinfo {author} {\bibfnamefont {S.}~\bibnamefont
  {Badoux}}, \bibinfo {author} {\bibfnamefont {J.}~\bibnamefont {Day}},
  \bibinfo {author} {\bibfnamefont {D.~A.}\ \bibnamefont {Bonn}}, \bibinfo
  {author} {\bibfnamefont {W.~N.}\ \bibnamefont {Hardy}}, \bibinfo {author}
  {\bibfnamefont {R.}~\bibnamefont {Liang}}, \bibinfo {author} {\bibfnamefont
  {N.}~\bibnamefont {Doiron-Leyraud}}, \ and\ \bibinfo {author} {\bibfnamefont
  {L.}~\bibnamefont {Taillefer}},\ }\href {\doibase 10.1103/PhysRevB.92.224502}
  {\bibfield  {journal} {\bibinfo  {journal} {Phys. Rev. B}\ }\textbf {\bibinfo
  {volume} {92}},\ \bibinfo {pages} {224502} (\bibinfo {year}
  {2015})}\BibitemShut {NoStop}%
\bibitem [{\citenamefont {D.~Pelc}(2016)}]{Pelc2016}%
  \BibitemOpen
  \bibfield  {author} {\bibinfo {author} {\bibfnamefont {H.~J. G. S. H. B. .
  M.~P.}\ \bibnamefont {D.~Pelc}, \bibfnamefont {M.~Vučković}},\ }\href
  {\doibase 10.1038/ncomms12775 (2016)} {\bibfield  {journal} {\bibinfo
  {journal} {Nat. Commun.}\ }\textbf {\bibinfo {volume} {7}},\ \bibinfo {pages}
  {12775} (\bibinfo {year} {2016})}\BibitemShut {NoStop}%
\bibitem [{\citenamefont {R.~Daou}(2010)}]{Daou2010}%
  \BibitemOpen
  \bibfield  {author} {\bibinfo {author} {\bibfnamefont {D.~L. O. C.-C. F. L.
  N. D.-L. B. J. R. R. L. D. A. B. W. N. H. . L.~T.}\ \bibnamefont {R.~Daou},
  \bibfnamefont {J.~Chang}},\ }\href {\doibase doi.org/10.1038/nature08716}
  {\bibfield  {journal} {\bibinfo  {journal} {Nature}\ }\textbf {\bibinfo
  {volume} {463}},\ \bibinfo {pages} {519} (\bibinfo {year}
  {2010})}\BibitemShut {NoStop}%
\bibitem [{\citenamefont {Makoto~Hashimoto}(2014)}]{Hashimoto2014}%
  \BibitemOpen
  \bibfield  {author} {\bibinfo {author} {\bibfnamefont {R.-H. H.-T. P. D. .
  Z.-X.~S.}\ \bibnamefont {Makoto~Hashimoto}, \bibfnamefont {Inna M.~Vishik}},\
  }\href {\doibase https://doi.org/10.1038/nphys3009} {\bibfield  {journal}
  {\bibinfo  {journal} {Nature Physics}\ }\textbf {\bibinfo {volume} {10}},\
  \bibinfo {pages} {483} (\bibinfo {year} {2014})}\BibitemShut {NoStop}%
\bibitem [{\citenamefont {Yamase}\ and\ \citenamefont
  {Kohno}(2000)}]{Yamase2000}%
  \BibitemOpen
  \bibfield  {author} {\bibinfo {author} {\bibfnamefont {H.}~\bibnamefont
  {Yamase}}\ and\ \bibinfo {author} {\bibfnamefont {H.}~\bibnamefont {Kohno}},\
  }\href {\doibase 10.1143/JPSJ.69.2151} {\bibfield  {journal} {\bibinfo
  {journal} {Journal of the Physical Society of Japan}\ }\textbf {\bibinfo
  {volume} {69}},\ \bibinfo {pages} {2151} (\bibinfo {year} {2000})},\ \Eprint
  {http://arxiv.org/abs/https://doi.org/10.1143/JPSJ.69.2151}
  {https://doi.org/10.1143/JPSJ.69.2151} \BibitemShut {NoStop}%
\bibitem [{\citenamefont {Okamoto}\ \emph {et~al.}(2010)\citenamefont
  {Okamoto}, \citenamefont {S\'en\'echal}, \citenamefont {Civelli},\ and\
  \citenamefont {Tremblay}}]{Okamoto2010}%
  \BibitemOpen
  \bibfield  {author} {\bibinfo {author} {\bibfnamefont {S.}~\bibnamefont
  {Okamoto}}, \bibinfo {author} {\bibfnamefont {D.}~\bibnamefont
  {S\'en\'echal}}, \bibinfo {author} {\bibfnamefont {M.}~\bibnamefont
  {Civelli}}, \ and\ \bibinfo {author} {\bibfnamefont {A.-M.~S.}\ \bibnamefont
  {Tremblay}},\ }\href {\doibase 10.1103/PhysRevB.82.180511} {\bibfield
  {journal} {\bibinfo  {journal} {Phys. Rev. B}\ }\textbf {\bibinfo {volume}
  {82}},\ \bibinfo {pages} {180511} (\bibinfo {year} {2010})}\BibitemShut
  {NoStop}%
\bibitem [{\citenamefont {Yamase}\ and\ \citenamefont
  {Metzner}(2007)}]{Yamase2007}%
  \BibitemOpen
  \bibfield  {author} {\bibinfo {author} {\bibfnamefont {H.}~\bibnamefont
  {Yamase}}\ and\ \bibinfo {author} {\bibfnamefont {W.}~\bibnamefont
  {Metzner}},\ }\href {\doibase 10.1103/PhysRevB.75.155117} {\bibfield
  {journal} {\bibinfo  {journal} {Phys. Rev. B}\ }\textbf {\bibinfo {volume}
  {75}},\ \bibinfo {pages} {155117} (\bibinfo {year} {2007})}\BibitemShut
  {NoStop}%
\bibitem [{\citenamefont {Kaczmarczyk}\ \emph {et~al.}(2016)\citenamefont
  {Kaczmarczyk}, \citenamefont {Schickling},\ and\ \citenamefont
  {B\"unemann}}]{Kaczmarczyk2016}%
  \BibitemOpen
  \bibfield  {author} {\bibinfo {author} {\bibfnamefont {J.}~\bibnamefont
  {Kaczmarczyk}}, \bibinfo {author} {\bibfnamefont {T.}~\bibnamefont
  {Schickling}}, \ and\ \bibinfo {author} {\bibfnamefont {J.}~\bibnamefont
  {B\"unemann}},\ }\href {\doibase 10.1103/PhysRevB.94.085152} {\bibfield
  {journal} {\bibinfo  {journal} {Phys. Rev. B}\ }\textbf {\bibinfo {volume}
  {94}},\ \bibinfo {pages} {085152} (\bibinfo {year} {2016})}\BibitemShut
  {NoStop}%
\bibitem [{\citenamefont {Kitatani}\ \emph {et~al.}(2017)\citenamefont
  {Kitatani}, \citenamefont {Tsuji},\ and\ \citenamefont
  {Aoki}}]{Kitatani2017}%
  \BibitemOpen
  \bibfield  {author} {\bibinfo {author} {\bibfnamefont {M.}~\bibnamefont
  {Kitatani}}, \bibinfo {author} {\bibfnamefont {N.}~\bibnamefont {Tsuji}}, \
  and\ \bibinfo {author} {\bibfnamefont {H.}~\bibnamefont {Aoki}},\ }\href
  {\doibase 10.1103/PhysRevB.95.075109} {\bibfield  {journal} {\bibinfo
  {journal} {Phys. Rev. B}\ }\textbf {\bibinfo {volume} {95}},\ \bibinfo
  {pages} {075109} (\bibinfo {year} {2017})}\BibitemShut {NoStop}%
\bibitem [{\citenamefont {Zegrodnik}\ and\ \citenamefont
  {Spa{\l}ek}(2018)}]{Zegrodnik2018}%
  \BibitemOpen
  \bibfield  {author} {\bibinfo {author} {\bibfnamefont {M.}~\bibnamefont
  {Zegrodnik}}\ and\ \bibinfo {author} {\bibfnamefont {J.}~\bibnamefont
  {Spa{\l}ek}},\ }\href {\doibase 10.1088/1367-2630/aac6f7} {\bibfield
  {journal} {\bibinfo  {journal} {New Journal of Physics}\ }\textbf {\bibinfo
  {volume} {20}},\ \bibinfo {pages} {063015} (\bibinfo {year}
  {2018})}\BibitemShut {NoStop}%
\bibitem [{\citenamefont {Slizovskiy}\ \emph {et~al.}(2018)\citenamefont
  {Slizovskiy}, \citenamefont {Rodriguez-Lopez},\ and\ \citenamefont
  {Betouras}}]{Slizovskiy2018}%
  \BibitemOpen
  \bibfield  {author} {\bibinfo {author} {\bibfnamefont {S.}~\bibnamefont
  {Slizovskiy}}, \bibinfo {author} {\bibfnamefont {P.}~\bibnamefont
  {Rodriguez-Lopez}}, \ and\ \bibinfo {author} {\bibfnamefont {J.~J.}\
  \bibnamefont {Betouras}},\ }\href {\doibase 10.1103/PhysRevB.98.075126}
  {\bibfield  {journal} {\bibinfo  {journal} {Phys. Rev. B}\ }\textbf {\bibinfo
  {volume} {98}},\ \bibinfo {pages} {075126} (\bibinfo {year}
  {2018})}\BibitemShut {NoStop}%
\bibitem [{\citenamefont {Kivelson}\ \emph {et~al.}(2004)\citenamefont
  {Kivelson}, \citenamefont {Fradkin},\ and\ \citenamefont
  {Geballe}}]{Kivelson2004}%
  \BibitemOpen
  \bibfield  {author} {\bibinfo {author} {\bibfnamefont {S.~A.}\ \bibnamefont
  {Kivelson}}, \bibinfo {author} {\bibfnamefont {E.}~\bibnamefont {Fradkin}}, \
  and\ \bibinfo {author} {\bibfnamefont {T.~H.}\ \bibnamefont {Geballe}},\
  }\href {\doibase 10.1103/PhysRevB.69.144505} {\bibfield  {journal} {\bibinfo
  {journal} {Phys. Rev. B}\ }\textbf {\bibinfo {volume} {69}},\ \bibinfo
  {pages} {144505} (\bibinfo {year} {2004})}\BibitemShut {NoStop}%
\bibitem [{\citenamefont {Fischer}\ and\ \citenamefont
  {Kim}(2011)}]{Fischer2011}%
  \BibitemOpen
  \bibfield  {author} {\bibinfo {author} {\bibfnamefont {M.~H.}\ \bibnamefont
  {Fischer}}\ and\ \bibinfo {author} {\bibfnamefont {E.-A.}\ \bibnamefont
  {Kim}},\ }\href {\doibase 10.1103/PhysRevB.84.144502} {\bibfield  {journal}
  {\bibinfo  {journal} {Phys. Rev. B}\ }\textbf {\bibinfo {volume} {84}},\
  \bibinfo {pages} {144502} (\bibinfo {year} {2011})}\BibitemShut {NoStop}%
\bibitem [{\citenamefont {Bulut}\ \emph {et~al.}(2013)\citenamefont {Bulut},
  \citenamefont {Atkinson},\ and\ \citenamefont {Kampf}}]{Bulut2013}%
  \BibitemOpen
  \bibfield  {author} {\bibinfo {author} {\bibfnamefont {S.}~\bibnamefont
  {Bulut}}, \bibinfo {author} {\bibfnamefont {W.~A.}\ \bibnamefont {Atkinson}},
  \ and\ \bibinfo {author} {\bibfnamefont {A.~P.}\ \bibnamefont {Kampf}},\
  }\href {\doibase 10.1103/PhysRevB.88.155132} {\bibfield  {journal} {\bibinfo
  {journal} {Phys. Rev. B}\ }\textbf {\bibinfo {volume} {88}},\ \bibinfo
  {pages} {155132} (\bibinfo {year} {2013})}\BibitemShut {NoStop}%
\bibitem [{\citenamefont {Tsuchiizu}\ \emph {et~al.}(2018)\citenamefont
  {Tsuchiizu}, \citenamefont {Kawaguchi}, \citenamefont {Yamakawa},\ and\
  \citenamefont {Kontani}}]{Tsuchiizu2018}%
  \BibitemOpen
  \bibfield  {author} {\bibinfo {author} {\bibfnamefont {M.}~\bibnamefont
  {Tsuchiizu}}, \bibinfo {author} {\bibfnamefont {K.}~\bibnamefont
  {Kawaguchi}}, \bibinfo {author} {\bibfnamefont {Y.}~\bibnamefont {Yamakawa}},
  \ and\ \bibinfo {author} {\bibfnamefont {H.}~\bibnamefont {Kontani}},\ }\href
  {\doibase 10.1103/PhysRevB.97.165131} {\bibfield  {journal} {\bibinfo
  {journal} {Phys. Rev. B}\ }\textbf {\bibinfo {volume} {97}},\ \bibinfo
  {pages} {165131} (\bibinfo {year} {2018})}\BibitemShut {NoStop}%
\bibitem [{\citenamefont {Zegrodnik}\ and\ \citenamefont
  {Spa\l{}ek}(2017)}]{Zegrodnik_1band_1}%
  \BibitemOpen
  \bibfield  {author} {\bibinfo {author} {\bibfnamefont {M.}~\bibnamefont
  {Zegrodnik}}\ and\ \bibinfo {author} {\bibfnamefont {J.}~\bibnamefont
  {Spa\l{}ek}},\ }\href {\doibase 10.1103/PhysRevB.96.054511} {\bibfield
  {journal} {\bibinfo  {journal} {Phys. Rev. B}\ }\textbf {\bibinfo {volume}
  {96}},\ \bibinfo {pages} {054511} (\bibinfo {year} {2017})}\BibitemShut
  {NoStop}%
\bibitem [{\citenamefont {Spa\l{}ek}\ \emph {et~al.}(2017)\citenamefont
  {Spa\l{}ek}, \citenamefont {Zegrodnik},\ and\ \citenamefont
  {Kaczmarczyk}}]{Zegrodnik_1band_2}%
  \BibitemOpen
  \bibfield  {author} {\bibinfo {author} {\bibfnamefont {J.}~\bibnamefont
  {Spa\l{}ek}}, \bibinfo {author} {\bibfnamefont {M.}~\bibnamefont
  {Zegrodnik}}, \ and\ \bibinfo {author} {\bibfnamefont {J.}~\bibnamefont
  {Kaczmarczyk}},\ }\href {\doibase 10.1103/PhysRevB.95.024506} {\bibfield
  {journal} {\bibinfo  {journal} {Phys. Rev. B}\ }\textbf {\bibinfo {volume}
  {95}},\ \bibinfo {pages} {024506} (\bibinfo {year} {2017})}\BibitemShut
  {NoStop}%
\bibitem [{\citenamefont {Zegrodnik}\ \emph {et~al.}(2019)\citenamefont
  {Zegrodnik}, \citenamefont {Biborski}, \citenamefont {Fidrysiak},\ and\
  \citenamefont {Spa\l{}ek}}]{Zegrodnik_3band}%
  \BibitemOpen
  \bibfield  {author} {\bibinfo {author} {\bibfnamefont {M.}~\bibnamefont
  {Zegrodnik}}, \bibinfo {author} {\bibfnamefont {A.}~\bibnamefont {Biborski}},
  \bibinfo {author} {\bibfnamefont {M.}~\bibnamefont {Fidrysiak}}, \ and\
  \bibinfo {author} {\bibfnamefont {J.}~\bibnamefont {Spa\l{}ek}},\ }\href
  {\doibase 10.1103/PhysRevB.99.104511} {\bibfield  {journal} {\bibinfo
  {journal} {Phys. Rev. B}\ }\textbf {\bibinfo {volume} {99}},\ \bibinfo
  {pages} {104511} (\bibinfo {year} {2019})}\BibitemShut {NoStop}%
\bibitem [{\citenamefont {Hybertsen}\ \emph {et~al.}(1989)\citenamefont
  {Hybertsen}, \citenamefont {Schl\"uter},\ and\ \citenamefont
  {Christensen}}]{3b_DFT_1989}%
  \BibitemOpen
  \bibfield  {author} {\bibinfo {author} {\bibfnamefont {M.~S.}\ \bibnamefont
  {Hybertsen}}, \bibinfo {author} {\bibfnamefont {M.}~\bibnamefont
  {Schl\"uter}}, \ and\ \bibinfo {author} {\bibfnamefont {N.~E.}\ \bibnamefont
  {Christensen}},\ }\href {\doibase 10.1103/PhysRevB.39.9028} {\bibfield
  {journal} {\bibinfo  {journal} {Phys. Rev. B}\ }\textbf {\bibinfo {volume}
  {39}},\ \bibinfo {pages} {9028} (\bibinfo {year} {1989})}\BibitemShut
  {NoStop}%
\bibitem [{\citenamefont {McMahan}\ \emph {et~al.}(1990)\citenamefont
  {McMahan}, \citenamefont {Annett},\ and\ \citenamefont
  {Martin}}]{3b_DFT_1990}%
  \BibitemOpen
  \bibfield  {author} {\bibinfo {author} {\bibfnamefont {A.~K.}\ \bibnamefont
  {McMahan}}, \bibinfo {author} {\bibfnamefont {J.~F.}\ \bibnamefont {Annett}},
  \ and\ \bibinfo {author} {\bibfnamefont {R.~M.}\ \bibnamefont {Martin}},\
  }\href {\doibase 10.1103/PhysRevB.42.6268} {\bibfield  {journal} {\bibinfo
  {journal} {Phys. Rev. B}\ }\textbf {\bibinfo {volume} {42}},\ \bibinfo
  {pages} {6268} (\bibinfo {year} {1990})}\BibitemShut {NoStop}%
\bibitem [{\citenamefont {Hirayama}\ \emph {et~al.}(2018)\citenamefont
  {Hirayama}, \citenamefont {Yamaji}, \citenamefont {Misawa},\ and\
  \citenamefont {Imada}}]{Hirayama_GWDFT_2018}%
  \BibitemOpen
  \bibfield  {author} {\bibinfo {author} {\bibfnamefont {M.}~\bibnamefont
  {Hirayama}}, \bibinfo {author} {\bibfnamefont {Y.}~\bibnamefont {Yamaji}},
  \bibinfo {author} {\bibfnamefont {T.}~\bibnamefont {Misawa}}, \ and\ \bibinfo
  {author} {\bibfnamefont {M.}~\bibnamefont {Imada}},\ }\href {\doibase
  10.1103/PhysRevB.98.134501} {\bibfield  {journal} {\bibinfo  {journal} {Phys.
  Rev. B}\ }\textbf {\bibinfo {volume} {98}},\ \bibinfo {pages} {134501}
  (\bibinfo {year} {2018})}\BibitemShut {NoStop}%
\bibitem [{\citenamefont {Kaczmarczyk}\ \emph {et~al.}(2013)\citenamefont
  {Kaczmarczyk}, \citenamefont {Spa\l{}ek}, \citenamefont {Schickling},\ and\
  \citenamefont {B\"unemann}}]{Kaczmarczyk_2013}%
  \BibitemOpen
  \bibfield  {author} {\bibinfo {author} {\bibfnamefont {J.}~\bibnamefont
  {Kaczmarczyk}}, \bibinfo {author} {\bibfnamefont {J.}~\bibnamefont
  {Spa\l{}ek}}, \bibinfo {author} {\bibfnamefont {T.}~\bibnamefont
  {Schickling}}, \ and\ \bibinfo {author} {\bibfnamefont {J.}~\bibnamefont
  {B\"unemann}},\ }\href {\doibase 10.1103/PhysRevB.88.115127} {\bibfield
  {journal} {\bibinfo  {journal} {Phys. Rev. B}\ }\textbf {\bibinfo {volume}
  {88}},\ \bibinfo {pages} {115127} (\bibinfo {year} {2013})}\BibitemShut
  {NoStop}%
\bibitem [{\citenamefont {Kaczmarczyk}\ \emph {et~al.}(2014)\citenamefont
  {Kaczmarczyk}, \citenamefont {Bünemann},\ and\ \citenamefont
  {Spałek}}]{Kaczmarczyk2014}%
  \BibitemOpen
  \bibfield  {author} {\bibinfo {author} {\bibfnamefont {J.}~\bibnamefont
  {Kaczmarczyk}}, \bibinfo {author} {\bibfnamefont {J.}~\bibnamefont
  {Bünemann}}, \ and\ \bibinfo {author} {\bibfnamefont {J.}~\bibnamefont
  {Spałek}},\ }\href {http://stacks.iop.org/1367-2630/16/i=7/a=073018}
  {\bibfield  {journal} {\bibinfo  {journal} {New Journal of Physics}\ }\textbf
  {\bibinfo {volume} {16}},\ \bibinfo {pages} {073018} (\bibinfo {year}
  {2014})}\BibitemShut {NoStop}%
\bibitem [{\citenamefont {Wysoki\ifmmode~\acute{n}\else \'{n}\fi{}ski}\ \emph
  {et~al.}(2016)\citenamefont {Wysoki\ifmmode~\acute{n}\else \'{n}\fi{}ski},
  \citenamefont {Kaczmarczyk},\ and\ \citenamefont
  {Spa\l{}ek}}]{MWysokinski_Anderson}%
  \BibitemOpen
  \bibfield  {author} {\bibinfo {author} {\bibfnamefont {M.~M.}\ \bibnamefont
  {Wysoki\ifmmode~\acute{n}\else \'{n}\fi{}ski}}, \bibinfo {author}
  {\bibfnamefont {J.}~\bibnamefont {Kaczmarczyk}}, \ and\ \bibinfo {author}
  {\bibfnamefont {J.}~\bibnamefont {Spa\l{}ek}},\ }\href {\doibase
  10.1103/PhysRevB.94.024517} {\bibfield  {journal} {\bibinfo  {journal} {Phys.
  Rev. B}\ }\textbf {\bibinfo {volume} {94}},\ \bibinfo {pages} {024517}
  (\bibinfo {year} {2016})}\BibitemShut {NoStop}%
\bibitem [{\citenamefont {M\"unster}\ and\ \citenamefont
  {B\"unemann}(2016)}]{2band_Hub_Bunemann}%
  \BibitemOpen
  \bibfield  {author} {\bibinfo {author} {\bibfnamefont {K.~z.}\ \bibnamefont
  {M\"unster}}\ and\ \bibinfo {author} {\bibfnamefont {J.}~\bibnamefont
  {B\"unemann}},\ }\href {\doibase 10.1103/PhysRevB.94.045135} {\bibfield
  {journal} {\bibinfo  {journal} {Phys. Rev. B}\ }\textbf {\bibinfo {volume}
  {94}},\ \bibinfo {pages} {045135} (\bibinfo {year} {2016})}\BibitemShut
  {NoStop}%
\bibitem [{\citenamefont {Lyons}\ \emph {et~al.}(1988)\citenamefont {Lyons},
  \citenamefont {Fleury}, \citenamefont {Schneemeyer},\ and\ \citenamefont
  {Waszczak}}]{Lyons1988}%
  \BibitemOpen
  \bibfield  {author} {\bibinfo {author} {\bibfnamefont {K.~B.}\ \bibnamefont
  {Lyons}}, \bibinfo {author} {\bibfnamefont {P.~A.}\ \bibnamefont {Fleury}},
  \bibinfo {author} {\bibfnamefont {L.~F.}\ \bibnamefont {Schneemeyer}}, \ and\
  \bibinfo {author} {\bibfnamefont {J.~V.}\ \bibnamefont {Waszczak}},\ }\href
  {\doibase 10.1103/PhysRevLett.60.732} {\bibfield  {journal} {\bibinfo
  {journal} {Phys. Rev. Lett.}\ }\textbf {\bibinfo {volume} {60}},\ \bibinfo
  {pages} {732} (\bibinfo {year} {1988})}\BibitemShut {NoStop}%
\bibitem [{\citenamefont {Sugai}\ \emph {et~al.}(1988)\citenamefont {Sugai},
  \citenamefont {Shamoto},\ and\ \citenamefont {Sato}}]{Sugai1988}%
  \BibitemOpen
  \bibfield  {author} {\bibinfo {author} {\bibfnamefont {S.}~\bibnamefont
  {Sugai}}, \bibinfo {author} {\bibfnamefont {S.-i.}\ \bibnamefont {Shamoto}},
  \ and\ \bibinfo {author} {\bibfnamefont {M.}~\bibnamefont {Sato}},\ }\href
  {\doibase 10.1103/PhysRevB.38.6436} {\bibfield  {journal} {\bibinfo
  {journal} {Phys. Rev. B}\ }\textbf {\bibinfo {volume} {38}},\ \bibinfo
  {pages} {6436} (\bibinfo {year} {1988})}\BibitemShut {NoStop}%
\bibitem [{\citenamefont {Blumberg}\ \emph {et~al.}(1996)\citenamefont
  {Blumberg}, \citenamefont {Abbamonte}, \citenamefont {Klein}, \citenamefont
  {Lee}, \citenamefont {Ginsberg}, \citenamefont {Miller},\ and\ \citenamefont
  {Zibold}}]{Blumberg1996}%
  \BibitemOpen
  \bibfield  {author} {\bibinfo {author} {\bibfnamefont {G.}~\bibnamefont
  {Blumberg}}, \bibinfo {author} {\bibfnamefont {P.}~\bibnamefont {Abbamonte}},
  \bibinfo {author} {\bibfnamefont {M.~V.}\ \bibnamefont {Klein}}, \bibinfo
  {author} {\bibfnamefont {W.~C.}\ \bibnamefont {Lee}}, \bibinfo {author}
  {\bibfnamefont {D.~M.}\ \bibnamefont {Ginsberg}}, \bibinfo {author}
  {\bibfnamefont {L.~L.}\ \bibnamefont {Miller}}, \ and\ \bibinfo {author}
  {\bibfnamefont {A.}~\bibnamefont {Zibold}},\ }\href {\doibase
  10.1103/PhysRevB.53.R11930} {\bibfield  {journal} {\bibinfo  {journal} {Phys.
  Rev. B}\ }\textbf {\bibinfo {volume} {53}},\ \bibinfo {pages} {R11930}
  (\bibinfo {year} {1996})}\BibitemShut {NoStop}%
\bibitem [{\citenamefont {Mizuno}\ \emph {et~al.}(1998)\citenamefont {Mizuno},
  \citenamefont {Tohyama},\ and\ \citenamefont {Maekawa}}]{Mizuno1998}%
  \BibitemOpen
  \bibfield  {author} {\bibinfo {author} {\bibfnamefont {Y.}~\bibnamefont
  {Mizuno}}, \bibinfo {author} {\bibfnamefont {T.}~\bibnamefont {Tohyama}}, \
  and\ \bibinfo {author} {\bibfnamefont {S.}~\bibnamefont {Maekawa}},\ }\href
  {\doibase 10.1103/PhysRevB.58.R14713} {\bibfield  {journal} {\bibinfo
  {journal} {Phys. Rev. B}\ }\textbf {\bibinfo {volume} {58}},\ \bibinfo
  {pages} {R14713} (\bibinfo {year} {1998})}\BibitemShut {NoStop}%
\bibitem [{\citenamefont {Ogata}\ and\ \citenamefont
  {Fukuyama}(2008)}]{tJmodel_rev_2008}%
  \BibitemOpen
  \bibfield  {author} {\bibinfo {author} {\bibfnamefont {M.}~\bibnamefont
  {Ogata}}\ and\ \bibinfo {author} {\bibfnamefont {H.}~\bibnamefont
  {Fukuyama}},\ }\href {http://stacks.iop.org/0034-4885/71/i=3/a=036501}
  {\bibfield  {journal} {\bibinfo  {journal} {Reports on Progress in Physics}\
  }\textbf {\bibinfo {volume} {71}},\ \bibinfo {pages} {036501} (\bibinfo
  {year} {2008})}\BibitemShut {NoStop}%
\end{thebibliography}%

\end{document}